\documentclass[12pt]{article}
\usepackage{ragged2e}
\usepackage{setspace} 
\usepackage{multicol}
\usepackage{multirow}
\usepackage{graphicx}
\usepackage{tikz}
\usepackage{float}
\usetikzlibrary{automata,positioning}
\usepackage{longtable}
\usepackage[margin=1in]{geometry}
\usepackage{tikz}
\usepackage{indentfirst}

\begin{document}

        \centering

\vspace*{.25cm}
    {\LARGE\bfseries Measuring Decidability as Related to Busy Beaver Numbers\par}
\vspace{.65cm}

\begin{minipage}{0.45\textwidth}
        \centering
{\small

    Gurpreet Tandi\\
    Bakersfield College\\
    tandi.sgurpreet@gmail.com

}
\end{minipage}
\hspace{-1cm}
\begin{minipage}{0.45\textwidth}
        \centering
{\small

    Josue Gonzalez-Hendrix\\
    Bakersfield College\\
    jgonzalezhendrix@gmail.com

}

\end{minipage}

    \vspace{0.5cm}

\begin{minipage}{0.45\textwidth}
        \centering
{\small

    Dr. Jonathan P. Brown, PhD\\
    Bakersfield College\\
    jonathan.brown@bakersfieldcollege.edu

}

\end{minipage}

    \vspace{0.5cm}

{\large August 11, 2026}

    \vspace{1cm}

\begin{minipage}{0.875\textwidth}
\centering
\textbf{Abstract}\\[0.5em]
\begin{minipage}{\textwidth}
        \justifying
\small    
    The theoretical existence of Busy Beaver numbers provides a new notion for decidability and corresponding heuristics for conjectures. The minimum number of states in which a conjecture can be modeled gives a classification of what logic system can describe said conjecture. In this work, we construct explicit Turing machines that search for a solution to Brocard's problem greater than $n = 7$ and a Fermat prime beyond $n = 4$ which halt if and only if such solutions exist.
\end{minipage}

\end{minipage}
\begin{flushleft}
\section{Introduction}
\subsection{Background and Motivation}

\justifying
    \noindent Brocard's problem was first posed in 1876 and the hunt for Fermat primes began in around 1650\cite{Brocard1876, Euler1738}. The decidability of whether or not unknown solutions exist for Brocard's problem is unknown along with the decidability of wether or not unknown Fermat primes exist. In this paper, we provide upper bounds for the number of states TMs (TM) searching for such numbers can be composed of by describing a TM which halts if and only if it finds an unknown solution to Brocard's problem and another which halts if and only if it finds an unknown Fermat prime.

    There exist two widely accepted notions of decidability: a problem is decidable if a TM that always halts in finite time can be constructed which returns a yes or no for a given input (known as `TM decidable'), or if there exists an algorithmic procedure for determining wether a statement is true or false in finitely many steps.

    We may use Busy Beaver numbers to circumvent the requirement of a TM always halting if the TM is designed to halt if and only if a statement is true which means that if the TM does not halt then the statement is false. Furthermore, we associate TMs with logic systems, by the number of states a TM contains corresponding to the strength of the logic system. That is, if the smallest TM which describes a conjecture is smaller than one which describes a different conjecture, the first conjecture requires a weaker logic system. In this way, we may describe the conjecture requiring a weaker logic system as having a higher decidability than the other conjecture.

\subsection{Turing Machines}

    {\noindent While there are many different types of TMs, which can emulate one another, we use the definition of a TM relevant to busy beaver numbers\cite{Rado1962}. In this definition, there is one tape composed of an infinite number in both directions of sequential cells which initially contain zeroes and may contain either a one or a zero, and states read the current value of the current cell, write, move to an adjacent cell, then transition to another state, with the latter three actions depending on which value was read.}

\subsection{The Busy Beaver Function}

    {\noindent The busy beaver number of $N$, (BB($N$)), describes the maximum number state transitions a TM of size N that halts can take before halting\cite{Rado1962}. For example, if a TM T is composed of four states, and runs for at least $108$ steps, we know it will never halt because BB($4$) = 107\cite{Brady1983}.
    
    The function BB($N$) is well defined.  To see this, consider the set of TM of size $N$ which always halt.   There are finitely many TMs of size $N$, and so this subset is also finite.  BB($N$) is just the maximum of steps required to halt on this set.

    The Busy Beaver function is proven undecidable, otherwise the halting problem would be decidable \cite{Rado1962}.   This makes the the BB based notion of decidability a weaker notion than standard decidability.  However, this loosening of conditions does induce an interesting heuristic for decidability that allows for categorization of conjectures.
    
\section{A TM That Halts If and Only If It Finds an Unknown Fermat Prime}

\subsection{Description}

    \noindent The Fermat number of $n$, $F_n$, is $2^{2^n}$. Since around the 1650's, we have known these to be prime for $n = 0, 1, 2, 3,  4$, but not any other numbers\cite{Euler1738,CrandallPomerance}.  It has been a long-standing conjecture that these are the only Fermat Primes.
    
    This TM has three sections, described in appendix A, B, and C. The first section computes $2^{16}$ given a tape populated only by zeros, the second section squares the number on the tape and adds one to it, and the third section halts if the number on the tape is prime and decrements the number by one if it is composite and calls the second section of the machine. The numbers recorded on the tape are always in unary. By combining states which will only ever read a one or a zero into states which have behaviors for both read cases, we find that the upper bound for the number of states a TM which halts if and only if at least one unknown Fermat prime exists is 72 states. Furthermore, states labeled ``NM'' indicate the transition to the first state of the next section being called.

\subsection{Explanation}

\subsubsection{Initialization}

    \noindent The first section of the machine computes the largest power of two corresponding to a known Fermat prime, the fourth Fermat number minus $1$.
\renewcommand{\arraystretch}{1.4}
\begin{center}
\begin{tabular}{|c|c|c|c|c|c|c|c|}
\hline
0 & 0 & 0 & 0 & 0 & 0 & 0 & 0 \\
\hline
\end{tabular}
$\rightarrow$
\renewcommand{\arraystretch}{1.4} 
\begin{tabular}{|c|c|c|c|c|c|c|c|}
\hline
1 & 1 & 1 & 0 & 1 & 1 & 1 & 1 \\
\hline
\end{tabular}
\end{center}

States 25-34 on Table 2 write four and three in unary onto the blank tape, leaving the current cell as the leftmost cell shown above. 

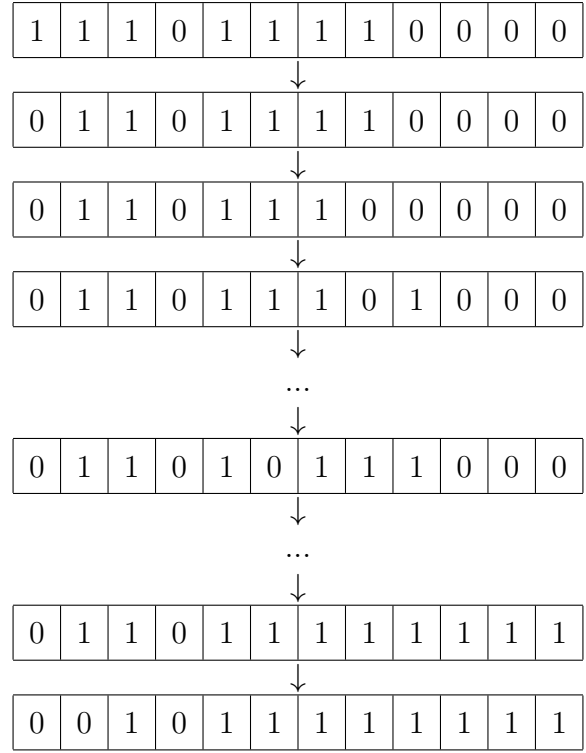
\renewcommand{\arraystretch}{1.4}
\begin{center}
\begin{tabular}{|c|c|c|c|c|c|c|c|c|c|c|c|}
\hline
1 & 1 & 1 & 0 & 1 & 1 & 1 & 1 & 0 & 0 & 0 & 0 \\
\hline
\end{tabular}\\
$\downarrow$
\\
\renewcommand{\arraystretch}{1.4} 
\begin{tabular}{|c|c|c|c|c|c|c|c|c|c|c|c|}
\hline
0 & 1 & 1 & 0 & 1 & 1 & 1 & 1 & 0 & 0 & 0 & 0 \\
\hline
\end{tabular}
\\
$\downarrow$
\\
\renewcommand{\arraystretch}{1.4} 
\begin{tabular}{|c|c|c|c|c|c|c|c|c|c|c|c|}
\hline
0 & 1 & 1 & 0 & 1 & 1 & 1 & 0 & 0 & 0 & 0 & 0 \\
\hline
\end{tabular}
\\
$\downarrow$
\\
\renewcommand{\arraystretch}{1.4} 
\begin{tabular}{|c|c|c|c|c|c|c|c|c|c|c|c|}
\hline
0 & 1 & 1 & 0 & 1 & 1 & 1 & 0 & 1 & 0 & 0 & 0 \\
\hline
\end{tabular}
\\
$\downarrow$
\\
...
\\
$\downarrow$
\\
\renewcommand{\arraystretch}{1.4} 
\begin{tabular}{|c|c|c|c|c|c|c|c|c|c|c|c|}
\hline
0 & 1 & 1 & 0 & 1 & 0 & 1 & 1 & 1 & 0 & 0 & 0 \\
\hline
\end{tabular}
\\
$\downarrow$
\\
...
\\
$\downarrow$
\\
\renewcommand{\arraystretch}{1.4} 
\begin{tabular}{|c|c|c|c|c|c|c|c|c|c|c|c|}
\hline
0 & 1 & 1 & 0 & 1 & 1 & 1 & 1 & 1 & 1 & 1 & 1 \\
\hline
\end{tabular}
\\
$\downarrow$
\\
\renewcommand{\arraystretch}{1.4} 
\begin{tabular}{|c|c|c|c|c|c|c|c|c|c|c|c|}
\hline
0 & 0 & 1 & 0 & 1 & 1 & 1 & 1 & 1 & 1 & 1 & 1 \\
\hline
\end{tabular}
\end{center}

    Sates 2-11 double the unary four three times by decrementing the unary three as an index, as shown above. If we call the number being doubled $x$ and the index $i$, we can say that state 9 dictates whether $x$ will be doubled by checking if $i$ is equal to zero. When $i$ is equal to zero, $x$ is equal to $16$. State 0 and states 12-14 set $x$ equal to one and use the $16$ as the index, allowing states 15-23 to double $x$ $i$ times as states 2-11 did. When state 15 detects $i$ is equal to zero, the current cell is set to the leftmost cell of the current number and the next section of the TM is called.

\subsubsection{Calculating the Next Fermat Number}

        \noindent We obtain the next Fermat number by squaring one less than previous Fermat number, and then adding one. Let $x$ be the number being squared on the tape. States 0-4 on Table 2 skip a cell and copy $x$ to the right of itself, as shown below. 
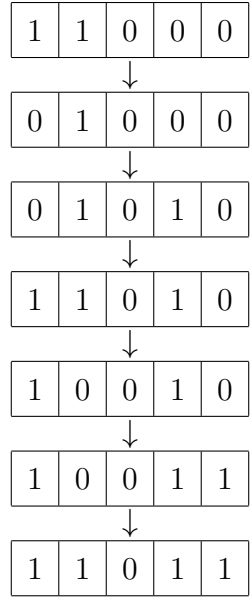
\begin{center}
   
\renewcommand{\arraystretch}{1.4}
\begin{tabular}{|c|c|c|c|c|}
\hline
1 & 1 & 0 & 0 & 0 \\
\hline
\end{tabular}
\\
$\downarrow$
\\
\renewcommand{\arraystretch}{1.4} 
\begin{tabular}{|c|c|c|c|c|}
\hline
0 & 1 & 0 & 0 & 0 \\
\hline
\end{tabular}\\
$\downarrow$
\\
\renewcommand{\arraystretch}{1.4} 
\begin{tabular}{|c|c|c|c|c|}
\hline
0 & 1 & 0 & 1 & 0 \\
\hline
\end{tabular}\\
$\downarrow$
\\
\renewcommand{\arraystretch}{1.4} 
\begin{tabular}{|c|c|c|c|c|}
\hline
1 & 1 & 0 & 1 & 0 \\
\hline
\end{tabular}\\
$\downarrow$
\\
\renewcommand{\arraystretch}{1.4} 
\begin{tabular}{|c|c|c|c|c|}
\hline
1 & 0 & 0 & 1 & 0 \\
\hline
\end{tabular}\\
$\downarrow$
\\
\renewcommand{\arraystretch}{1.4} 
\begin{tabular}{|c|c|c|c|c|}
\hline
1 & 0 & 0 & 1 & 1 \\
\hline
\end{tabular}\\
$\downarrow$
\\
\renewcommand{\arraystretch}{1.4} 
\begin{tabular}{|c|c|c|c|c|}
\hline
1 & 1 & 0 & 1 & 1 \\
\hline
\end{tabular}\\
\end{center}

        \indent Now, we may square by multiplying the two unary numbers on the tape. Let the left number be called $i$ and the right be called $x$. To obtain our desired output, $x^2$, we multiply $x$ by $i$ and write the output to the right of $x$ separated by a cell containing zero, as below.
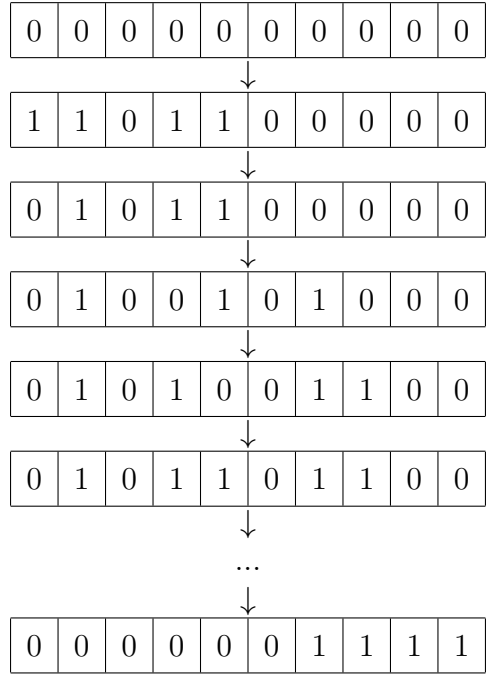
\begin{center}

\begin{tabular}{|c|c|c|c|c|c|c|c|c|c|}
\hline
0 & 0 & 0 & 0 & 0 & 0 & 0 & 0 & 0 & 0\\
\hline
\end{tabular}
\\
$\downarrow$
\\
\renewcommand{\arraystretch}{1.4} 
\begin{tabular}{|c|c|c|c|c|c|c|c|c|c|}
\hline
1 & 1 & 0 & 1 & 1 & 0 & 0 & 0 & 0 & 0\\
\hline
\end{tabular}\\
$\downarrow$
\\
\renewcommand{\arraystretch}{1.4} 
\begin{tabular}{|c|c|c|c|c|c|c|c|c|c|}
\hline
0 & 1 & 0 & 1 & 1 & 0 & 0 & 0 & 0 & 0\\
\hline
\end{tabular}\\
$\downarrow$
\\
\renewcommand{\arraystretch}{1.4} 
\begin{tabular}{|c|c|c|c|c|c|c|c|c|c|}
\hline
0 & 1 & 0 & 0 & 1 & 0 & 1 & 0 & 0 & 0\\
\hline
\end{tabular}\\
$\downarrow$
\\
\renewcommand{\arraystretch}{1.4} 
\begin{tabular}{|c|c|c|c|c|c|c|c|c|c|}
\hline
0 & 1 & 0 & 1 & 0 & 0 & 1 & 1 & 0 & 0\\
\hline
\end{tabular}\\
$\downarrow$
\\
\renewcommand{\arraystretch}{1.4} 
\begin{tabular}{|c|c|c|c|c|c|c|c|c|c|}
\hline
0 & 1 & 0 & 1 & 1 & 0 & 1 & 1 & 0 & 0\\
\hline
\end{tabular}\\
$\downarrow$\\
...\\
$\downarrow$
\\
\renewcommand{\arraystretch}{1.4} 
\begin{tabular}{|c|c|c|c|c|c|c|c|c|c|}
\hline
0 & 0 & 0 & 0 & 0 & 0 & 1 & 1 & 1 & 1\\
\hline
\end{tabular}\\

\end{center}

    States 6-15 on Table 2 erase the index and output $x$'s square to the right of $x$. State 17 on Table 2 must exist in order to traverse cells, and it erases $x$ and increments the square by one. Then, we have $F_n$ and the next section of the TM is called.

\subsubsection{Primality Check}

        \noindent We now have $F_n$ and must check for primality. If and only if $F_n$ is prime, the machine halts. If the number is composite, it will be decremented by one and the first state of Table 2 is called.
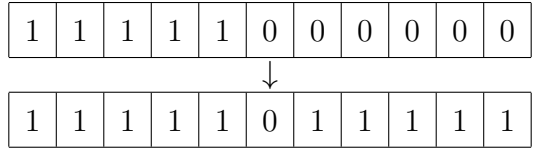
\begin{center}
\begin{tabular}{|c|c|c|c|c|c|c|c|c|c|c|}
\hline
1 & 1 & 1 & 1 & 1 & 0 & 0 & 0 & 0 & 0 & 0\\
\hline
\end{tabular}
\\
$\downarrow$
\\
\renewcommand{\arraystretch}{1.4} 
\begin{tabular}{|c|c|c|c|c|c|c|c|c|c|c|}
\hline
1 & 1 & 1 & 1 & 1 & 0 & 1 & 1 & 1 & 1 & 1\\
\hline
\end{tabular}\\
\end{center}

        Using an arbitrary number to demonstrate this algorithm, let's pretend the number we ended up with was five and call it $x$. States zero to four are a simple algorithm for duplicating $x$ like above.
        
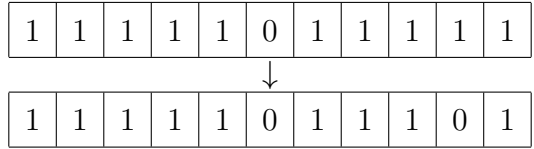
\begin{center}
    
\begin{tabular}{|c|c|c|c|c|c|c|c|c|c|c|}
\hline
1 & 1 & 1 & 1 & 1 & 0 & 1 & 1 & 1 & 1 & 1\\
\hline
\end{tabular}
\\
$\downarrow$
\\
\renewcommand{\arraystretch}{1.4} 
\begin{tabular}{|c|c|c|c|c|c|c|c|c|c|c|}
\hline
1 & 1 & 1 & 1 & 1 & 0 & 1 & 1 & 1 & 0 & 1\\
\hline
\end{tabular}\\
\end{center}

        States 5-8 separate the duplicate of $x$ into two indices: $i$, which is in the middle, and$r$, which is on the far right. This portion of the machine performs $x \%i$, decrementing $r$ and incrementing $i$ each time the module is not zero. When $r = 0$, the number is found to be prime. Note that $i$ must be initialized as $3$ for the machine to work properly and that this algorithm cannot test a number smaller than five.

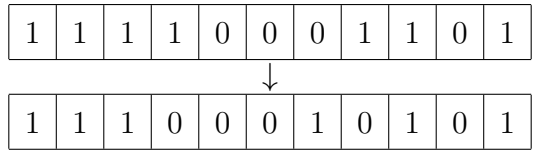
\begin{center}
\begin{tabular}{|c|c|c|c|c|c|c|c|c|c|c|}
\hline
1 & 1 & 1 & 1 & 0 & 0 & 0 & 1 & 1 & 0 & 1\\
\hline
\end{tabular}
\\
$\downarrow$
\\
\begin{tabular}{|c|c|c|c|c|c|c|c|c|c|c|}
\hline
1 & 1 & 1 & 0 & 0 & 0 & 1 & 0 & 1 & 0 & 1\\
\hline
\end{tabular}
\\
\end{center}

        States 10-16 and 32-34 increment the index and decrement $x$ by $3$. States 12, 16, and 35 handle the cases in which $x$ reaches zero, in which case $x$ is restored to its original size, $i$ is incremented and $r$ is decremented. Since the final element of $i$ being the index is a special case because if $x$ reaches zero at that point then the number is composite, the last element isn't switched to a zero and states 36-39 handle the logic of decrementing $x$ and deciding if the number is composite. If $x = 0$, state 38 calls state 43-46 and state 24 which repair the tape to solely include $x$ and decrement$x$ by one and call the first state of the square function.
        
        There are three separate cases for this method of factoring due to the fact that the index can be found at the beginning of $i$, the state immediately after the beginning of $i$, or a state after those two. The logic is based on the machine ``catching'' itself during the third case by checking if the number after the number being indexed is zero to accommodate for factoring by a number greater than three. State 35 also dictates whether or not this correction is made. Below is an example of a number being found to be prime, the only halt case.

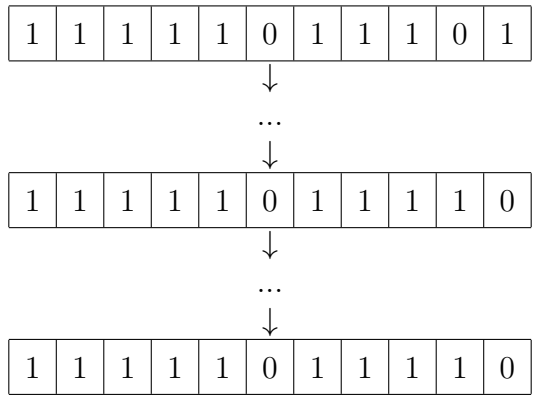
\begin{center}
\begin{tabular}{|c|c|c|c|c|c|c|c|c|c|c|}
\hline
1 & 1 & 1 & 1 & 1 & 0 & 1 & 1 & 1 & 0 & 1\\
\hline
\end{tabular}
\\
$\downarrow$
\\
...
\\
$\downarrow$
\\
\begin{tabular}{|c|c|c|c|c|c|c|c|c|c|c|}
\hline
1 & 1 & 1 & 1 & 1 & 0 & 1 & 1 & 1 & 1 & 0\\
\hline
\end{tabular}
\\
$\downarrow$
\\
...
\\
$\downarrow$
\\
\begin{tabular}{|c|c|c|c|c|c|c|c|c|c|c|}
\hline
1 & 1 & 1 & 1 & 1 & 0 & 1 & 1 & 1 & 1 & 0\\
\hline
\end{tabular}
\end{center}
       After $x$ has been restored, the machine attempts to increment $i$ and in doing so finds that $r = 0$, and therefore the machine halts. State specifically checks if $r = 0$. Below is an example of the machine finding a number composite.
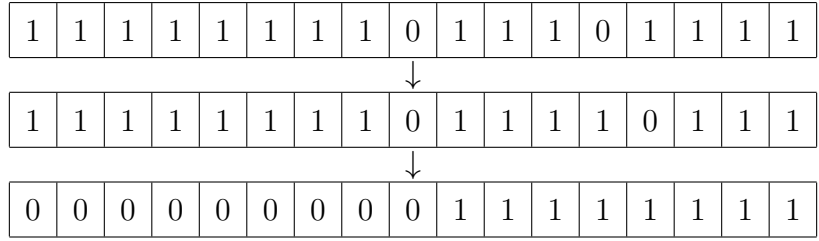
\begin{center}
    
\begin{tabular}{|c|c|c|c|c|c|c|c|c|c|c|c|c|c|c|c|c|}
\hline
1 & 1 & 1 & 1 & 1 & 1 & 1 & 1 & 0 & 1 & 1 & 1 & 0 & 1 & 1 & 1 & 1\\
\hline
\end{tabular}
\\
$\downarrow$
\\
\begin{tabular}{|c|c|c|c|c|c|c|c|c|c|c|c|c|c|c|c|c|}
\hline
1 & 1 & 1 & 1 & 1 & 1 & 1 & 1 & 0 & 1 & 1 & 1 & 1 & 0 & 1 & 1 & 1\\
\hline
\end{tabular}
\\
$\downarrow$
\\
\begin{tabular}{|c|c|c|c|c|c|c|c|c|c|c|c|c|c|c|c|c|}
\hline
0 & 0 & 0 & 0 & 0 & 0 & 0 & 0 & 0 & 1 & 1 & 1 & 1 & 1 & 1 & 1 & 1\\
\hline
\end{tabular}
\\
\end{center}
        The description of the nature of what the machine has done in this case is present earlier in the paper.\\
\section{A TM That Halts If and Only If It Finds an Unknown Solution to Brocard's Problem}

\subsection{Description}

\noindent Despite being explored since 1876, Brocard's problem $n!+1 = m^2$,  where $m$ and $n$ are integers, has no known solution for $n$ beyond $n = 4, 5, 7$ where m and n are integers\cite{Brocard1876,Ramanujan1913}.  It has long been conjectured that these are the only Brocard's numbers.
    
This TM two sections, denoted by `F' states and `S' states, both described in appendix D. The `F' states compute a factorial recursively using an index and then adds one to it and the `S' states check if the resulting number is a square by checking if it is the sum of consecutive odd numbers, in which case the machine halts. Combining the states which only have a 0 read behavior with the states that only have a $1$ read, we find that an upper bound for  the number of states a TM that halts if and only if at least one unknown solution to Brocard's problem exist is 43.

\subsection{Explanation}

\justifying
\noindent The machine starts off in the ``middle'' of an infinitely long tape. The states Df, Jf, Pf, Qf, and Es are used here initialize the tape with (from left to right) $n$ and $n!$ for $n=2$, separated by a 0, and leaves the tape to the leftmost 0 of $n$. States Af, Bf, Cf, and Df increment $n$, then enter $(n-1)!$.

\begin{figure}[ht!]
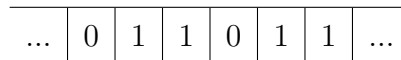

\centering
\begin{tabular}{c|c|c|c|c|c|c|c}
\hline 
... & 0 &1 & 1 & 0 & 1 & 1 & ... \\
\hline 
\end{tabular}

\caption{The tape after the first five steps}
\end{figure}

\justifying
States Ef, Ff, Gf, Hf If, and Jf adds $(n-1)*(n-1)!$ to a new sequence of $1$s to the  right of $(n-1)!$, separated by a $0$. Once it finishes, Hf, Kf, and Lf check to see for completion, transitioning to Af  if the calculation is finished. Once it has, the head returns to $n$, then  Mf, Nf, Of, Pf, Qf, Rf, and Sf  test if $n > 7$. If any of these states encounters a $1$,  Vf, Wf, Xf and Sf  restart the factorial calculation. If none of these states reads a $1$, Tf and Uf go to the right of $(n-1)!$ and $(n-1)*(n-1)!$, changing the separating $0$ to a $1$, leaving the value as $n!+1$, transitioning into  As, the beginning of the square checking half of the machine.

\begin{figure}[H]
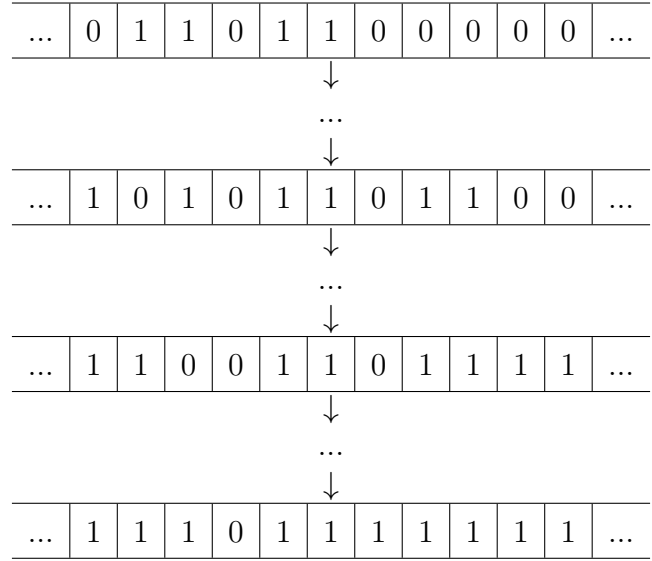

\centering

\begin{tabular}{c|c|c|c|c|c|c|c|c|c|c|c|c}
\hline 
...& 0 &1 & 1 & 0 & 1 & 1 & 0 & 0& 0 & 0 & 0 & ...\\
\hline 
\end{tabular}

$\downarrow$

...

$\downarrow$

\begin{tabular}{c|c|c|c|c|c|c|c|c|c|c|c|c}
\hline 
...& 1 &  0 & 1 & 0 & 1 & 1 & 0 & 1 & 1 & 0 & 0 & ...\\
\hline 
\end{tabular}

$\downarrow$

...

$\downarrow$

\begin{tabular}{c|c|c|c|c|c|c|c|c|c|c|c|c}
\hline 
...& 1 &  1 & 0 & 0 & 1 & 1 & 0 & 1 & 1 & 1 & 1 & ...\\
\hline 
\end{tabular}

$\downarrow$

...

$\downarrow$

\begin{tabular}{c|c|c|c|c|c|c|c|c|c|c|c|c}
\hline 
...& 1 &  1 & 1 & 0 & 1 & 1 & 1 & 1 & 1 & 1 & 1 & ...\\
\hline 
\end{tabular}

\caption{The tape during the calculation of 3!+1}
\end{figure}

\justifying
To check if $n!+1$ is a square, the machine recursively "subtracts" odd numbers, starting with $1$, from $n!+1$, replacing 1s with 0s. For an integer $2k+1$, it first subtracts $2k$ number of $0$s, only subtracting the final +1 if it hasn't subtracted away all of $n!+1$.

It keeps track of the current $k$ with a sequence of 1s to the right of $n!+1$.  Af changes the 0 to the right of $n!+1$ to a $1$  that will never be subtracted so it doesn't lose track of where the $n!+1$ is and goes right. Bs  goes to the right, leaving the 0 that will separate the $n!+1$ from $k$.


\begin{figure}[H]
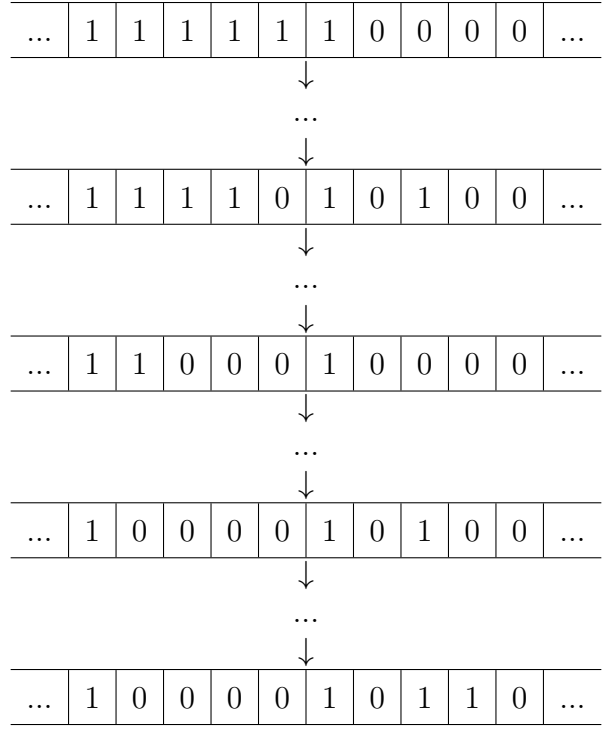

\centering

\begin{tabular}{c|c|c|c|c|c|c|c|c|c|c|c}
\hline
... & 1 & 1 & 1 & 1 & 1 & 1 & 0 & 0 & 0 & 0 &  ...\\
\hline
\end{tabular}

$\downarrow$

...

$\downarrow$

\begin{tabular}{c|c|c|c|c|c|c|c|c|c|c|c}
\hline
... & 1 & 1 & 1 & 1 & 0 & 1 & 0 & 1 & 0 & 0 &  ...\\
\hline
\end{tabular}

$\downarrow$

...

$\downarrow$

\begin{tabular}{c|c|c|c|c|c|c|c|c|c|c|c}
\hline
... & 1 & 1 & 0 & 0 & 0 & 1 & 0 & 0 & 0 & 0 &  ...\\
\hline
\end{tabular}

$\downarrow$

...

$\downarrow$

\begin{tabular}{c|c|c|c|c|c|c|c|c|c|c|c}
\hline
... & 1 & 0 & 0 & 0 & 0 & 1 & 0 & 1 & 0 & 0 &  ...\\
\hline
\end{tabular}

$\downarrow$

...

$\downarrow$

\begin{tabular}{c|c|c|c|c|c|c|c|c|c|c|c}
\hline
... & 1 & 0 & 0 & 0 & 0 & 1 & 0 & 1 & 1 & 0 &  ...\\
\hline
\end{tabular}

\caption{The tape while ``subtracting'' $2k+1$ from $n!+1$ for $k=0$, $k=1$, then assigning $k := 2$}
\end{figure}

Cs checks if it has finished subtracting $2k$, and if it hasn't, Ds, Es, Fs, Gs, Hs, Is, and Js subtract 2. If the machine reaches the end of $n!+1$ during this process, then Os, Ps, Qs, Rs, and Ss undo the subtraction, remove $k$, the separating $1$, and the right-most $1$ of $n!+1$ before transitioning into Kf to calculate the next factorial. If Cs didn't read a $0$, then the machine managed to finish subtracting $2k$, so Ls and Ms subtract the final $+1$. Ns checks if there is anything left over. If there isn't, then $n!+1$ is of the form \[\sum_{i=0}^{k}2k+1\] for some $k$ and is a square, so the machine halts. If there is still more, then Ks returns to $k$ and Bs increments it, allowing it to subtract the next odd integer.

    \vspace{.5cm}
    \begin{flushleft}{

\section{Related Works}

    {\begin{justify}Adam Yedida and Scott Aaronson's work \emph{A Relatively Small TM Whose Behavior Is Independent of Set Theory} inspired us in writing about TMs and logic systems by showing that it is possible to construct TMs which halt if and only if a logic system is inconsistent\cite{yedidia2016relativelysmallturingmachine}.\end{justify}}
    \vspace{.5cm}

    \vspace*{1cm}
\bibliographystyle{plain}
\bibliography{refernces}

    \vspace{.5cm}

    {\Large\textbf{Appendices}}
\\
    \vspace{.3cm}
    A~~~FP Table 1 and Visualization
    \vspace{.5cm}
\setlength{\LTpre}{0pt}
\setlength{\LTpost}{0pt}

\begin{longtable}{|l|l|l|l|l|}
\hline
\textbf{State} & \textbf{Reads} & \textbf{Writes} & \textbf{Moves} & \textbf{Calls} \\
\hline
0 & 0 & 1 & L & 13 \\
\hline
2 & 1 & 1 & R & 3 \\
\hline
2 & 0 & 0 & R & 2 \\
\hline
3 & 0 & 0 & L & 4 \\
\hline
3 & 1 & 1 & R & 3 \\
\hline
4 & 1 & 0 & R & 5 \\
\hline
5 & 0 & 1 & L & 6 \\
\hline
5 & 1 & 1 & R & 5 \\
\hline
6 & 0 & 1 & L & 7 \\
\hline
6 & 1 & 1 & L & 6 \\
\hline
7 & 0 & 0 & L & 11 \\
\hline
7 & 1 & 0 & R & 5 \\
\hline
9 & 0 & 0 & R & 12 \\
\hline
9 & 1 & 0 & R & 10 \\
\hline
10 & 0 & 0 & R & 2 \\
\hline
10 & 1 & 1 & R & 10 \\
\hline
11 & 0 & 0 & R & 9 \\
\hline
11 & 1 & 1 & L & 11 \\
\hline
12 & 0 & 0 & R & 0 \\
\hline
12 & 1 & 1 & R & 12 \\
\hline
13 & 0 & 0 & L & 14 \\
\hline
14 & 0 & 0 & R & 15 \\
\hline
14 & 1 & 1 & L & 14 \\
\hline
15 & 0 & 0 & R & NM \\
\hline
15 & 1 & 0 & R & 16 \\
\hline
16 & 0 & 0 & R & 17 \\
\hline
16 & 1 & 1 & R & 16 \\
\hline
17 & 1 & 1 & R & 18 \\
\hline
17 & 0 & 0 & R & 17 \\
\hline
18 & 1 & 1 & R & 18 \\
\hline
18 & 0 & 0 & L & 19 \\
\hline
19 & 1 & 0 & R & 20 \\
\hline
20 & 0 & 1 & L & 21 \\
\hline
20 & 1 & 1 & R & 20 \\
\hline
21 & 1 & 1 & L & 21 \\
\hline
21 & 0 & 1 & L & 22 \\
\hline
22 & 0 & 0 & L & 23 \\
\hline
22 & 1 & 0 & R & 20 \\
\hline
23 & 0 & 0 & R & 15 \\
\hline
23 & 1 & 1 & L & 23 \\
\hline
25 & 0 & 1 & L & 26 \\
\hline
26 & 0 & 1 & L & 27 \\
\hline
27 & 0 & 1 & L & 28 \\
\hline
28 & 0 & 1 & L & 29 \\
\hline
29 & 0 & 0 & L & 30 \\
\hline
30 & 0 & 1 & L & 31 \\
\hline
31 & 0 & 1 & L & 31 \\
\hline
34 & 0 & 0 & R & 9 \\
\hline
\end{longtable}

    \vspace{.3cm}
\begin{tikzpicture}[->,>=stealth,shorten >=1pt,auto,semithick,
    every node/.style={font=\tiny},
    xscale=.7,
    yscale=.7
]
\tikzstyle{every state}=[
    draw=black,
    thick,
    fill=white,
    minimum size=12pt,
    inner sep=1pt,
    font=\tiny
]
  \node[state] (0) at (28.05,-7.725) {q0};
  \node[state] (2) at (28.175,-3.325) {q2};
  \node[state] (3) at (30.8,-3.325) {q3};
  \node[state] (4) at (33.625,-3.325) {q4};
  \node[state] (5) at (36.3,-3.325) {q5};
  \node[state] (6) at (39.2,-3.325) {q6};
  \node[state] (7) at (36.3,-6.225) {q7};
  \node[state] (9) at (21.85,-3.325) {q9};
  \node[state] (10) at (25.3,-3.325) {q10};
  \node[state] (11) at (24.35,-6.2) {q11};
  \node[state] (12) at (21.85,-7.725) {q12};
  \node[state] (13) at (24.125,-10.4) {q13};
  \node[state] (14) at (21.875,-10.4) {q14};
  \node[state] (15) at (21.875,-13.55) {q15};
  \node[state] (16) at (24.975,-13.55) {q16};
  \node[state] (17) at (27.725,-13.55) {q17};
  \node[state] (18) at (30.2,-13.55) {q18};
  \node[state] (19) at (33.2,-13.55) {q19};
  \node[state] (20) at (35.725,-13.55) {q20};
  \node[state] (21) at (38.85,-13.55) {q21};
  \node[state] (22) at (35.725,-17.0) {q22};
  \node[state] (23) at (24.175,-17.0) {q23};
  \node[state,accepting] (24) at (21.875,-20.0) {NM};
  \node[state,initial] (25) at (16.7,-4.25) {q25};
  \node[state] (26) at (16.7,-7.275) {q26};
  \node[state] (27) at (16.7,-10.55) {q27};
  \node[state] (28) at (16.7,-13.55) {q28};
  \node[state] (29) at (16.7,-16.475) {q29};
  \node[state] (30) at (16.7,-19.8) {q30};
  \node[state] (31) at (19.725,-19.825) {q31};
  \node[state] (34) at (19.725,-11.325) {q34};
  \draw[->] (6) -- (7) node[midway, above] {\scriptsize 0, 1, L};
  \draw[->] (6) edge[loop above] node {\scriptsize 1, 1, L} (6);
  \draw[->] (23) -- (15) node[midway, above] {\scriptsize 0, 0, R};
  \draw[->] (23) edge[loop above] node {\scriptsize 1, 1, L} (23);
  \draw[->] (9) -- (12) node[midway, above] {\scriptsize 0, 0, R};
  \draw[->] (9) -- (10) node[midway, above] {\scriptsize 1, 0, R};
  \draw[->] (5) -- (6) node[midway, above] {\scriptsize 0, 1, L};
  \draw[->] (5) edge[loop above] node {\scriptsize 1, 1, R} (5);
  \draw[->] (17) -- (18) node[midway, above] {\scriptsize 1, 1, R};
  \draw[->] (17) edge[loop above] node {\scriptsize 0, 0, R} (17);
  \draw[->] (30) -- (31) node[midway, above] {\scriptsize 0, 1, L};
  \draw[->] (16) -- (17) node[midway, above] {\scriptsize 0, 0, R};
  \draw[->] (16) edge[loop above] node {\scriptsize 1, 1, R} (16);
  \draw[->] (34) -- (9) node[midway, above] {\scriptsize 0, 0, R};
  \draw[->] (2) -- (3) node[midway, above] {\scriptsize 1, 1, R};
  \draw[->] (2) edge[loop above] node {\scriptsize 0, 0, R} (2);
  \draw[->] (4) -- (5) node[midway, above] {\scriptsize 1, 0, R};
  \draw[->] (22) -- (23) node[midway, above] {\scriptsize 0, 0, L};
  \draw[->] (22) -- (20) node[midway, above] {\scriptsize 1, 0, R};
  \draw[->] (0) -- (13) node[midway, above] {\scriptsize 0, 1, L};
  \draw[->] (12) -- (0) node[midway, above] {\scriptsize 0, 0, R};
  \draw[->] (12) edge[loop above] node {\scriptsize 1, 1, R} (12);
  \draw[->] (27) -- (28) node[midway, above] {\scriptsize 0, 1, L};
  \draw[->] (3) -- (4) node[midway, above] {\scriptsize 0, 0, L};
  \draw[->] (3) edge[loop above] node {\scriptsize 1, 1, R} (3);
  \draw[->] (25) -- (26) node[midway, above] {\scriptsize 0, 1, L};
  \draw[->] (15) -- (24) node[midway, above] {\scriptsize 0, 0, R};
  \draw[->] (15) -- (16) node[midway, above] {\scriptsize 1, 0, R};
  \draw[->] (31) -- (34) node[midway, above] {\scriptsize 0, 1, L};
  \draw[->] (20) -- (21) node[midway, above] {\scriptsize 0, 1, L};
  \draw[->] (20) edge[loop above] node {\scriptsize 1, 1, R} (20);
  \draw[->] (14) -- (15) node[midway, above] {\scriptsize 0, 0, R};
  \draw[->] (14) edge[loop above] node {\scriptsize 1, 1, L} (14);
  \draw[->] (7) -- (11) node[midway, above] {\scriptsize 0, 0, L};
  \draw[->] (7) -- (5) node[midway, above] {\scriptsize 1, 0, R};
  \draw[->] (28) -- (29) node[midway, above] {\scriptsize 0, 1, L};
  \draw[->] (29) -- (30) node[midway, above] {\scriptsize 0, 0, L};
  \draw[->] (13) -- (14) node[midway, above] {\scriptsize 0, 0, L};
  \draw[->] (10) -- (2) node[midway, above] {\scriptsize 0, 0, R};
  \draw[->] (10) edge[loop above] node {\scriptsize 1, 1, R} (10);
  \draw[->] (11) -- (9) node[midway, above] {\scriptsize 0, 0, R};
  \draw[->] (11) edge[loop above] node {\scriptsize 1, 1, L} (11);
  \draw[->] (21) edge[loop above] node {\scriptsize 1, 1, L} (21);
  \draw[->] (21) -- (22) node[midway, above] {\scriptsize 0, 1, L};
  \draw[->] (26) -- (27) node[midway, above] {\scriptsize 0, 1, L};
  \draw[->] (19) -- (20) node[midway, above] {\scriptsize 1, 0, R};
  \draw[->] (18) edge[loop above] node {\scriptsize 1, 1, R} (18);
  \draw[->] (18) -- (19) node[midway, above] {\scriptsize 0, 0, L};
\end{tikzpicture}
    \vspace{.5cm}
    B~~~FP Table 2 and Visualization
    \vspace{.5cm}
\setlength{\LTpre}{0pt}
\setlength{\LTpost}{0pt}
\begin{longtable}{|l|l|l|l|l|}
\hline
\textbf{State} & \textbf{Reads} & \textbf{Writes} & \textbf{Moves} & \textbf{Calls} \\
\hline
0 & 0 & 0 & L & 5 \\
\hline
0 & 1 & 0 & R & 1 \\
\hline
1 & 1 & 1 & R & 1 \\
\hline
1 & 0 & 0 & R & 2 \\
\hline
2 & 0 & 1 & L & 3 \\
\hline
2 & 1 & 1 & R & 2 \\
\hline
3 & 0 & 0 & L & 4 \\
\hline
3 & 1 & 1 & L & 3 \\
\hline
4 & 0 & 1 & R & 0 \\
\hline
4 & 1 & 1 & L & 4 \\
\hline
5 & 0 & 0 & R & 6 \\
\hline
5 & 1 & 1 & L & 5 \\
\hline
6 & 0 & 0 & R & 17 \\
\hline
6 & 1 & 0 & R & 7 \\
\hline
7 & 0 & 0 & R & 8 \\
\hline
7 & 1 & 1 & R & 7 \\
\hline
8 & 1 & 0 & R & 9 \\
\hline
9 & 0 & 0 & R & 10 \\
\hline
9 & 1 & 1 & R & 9 \\
\hline
10 & 0 & 1 & L & 11 \\
\hline
10 & 1 & 1 & R & 10 \\
\hline
11 & 0 & 0 & L & 12 \\
\hline
11 & 1 & 1 & L & 11 \\
\hline
12 & 1 & 1 & L & 13 \\
\hline
12 & 0 & 1 & L & 14 \\
\hline
13 & 1 & 1 & L & 13 \\
\hline
13 & 0 & 1 & R & 8 \\
\hline
14 & 0 & 0 & L & 15 \\
\hline
14 & 1 & 1 & L & 14 \\
\hline
15 & 0 & 0 & R & 6 \\
\hline
15 & 1 & 1 & L & 15 \\
\hline
17 & 0 & 1 & R & NM \\
\hline
17 & 1 & 0 & R & 17 \\
\hline
\end{longtable}

    \vspace{.3cm}
\begin{tikzpicture}[->,>=stealth,shorten >=1pt,auto,semithick,
    every node/.style={font=\tiny},
    xscale=.55,
    yscale=.7
]
\tikzstyle{every state}=[
    draw=black,
    thick,
    fill=white,
    minimum size=12pt,
    inner sep=1pt,
    font=\tiny
]
  \node[state,initial] (0) at (29.125,-3.425) {q0};
  \node[state] (1) at (32.375,-5.325) {q1};
  \node[state] (2) at (34.775,-5.325) {q2};
  \node[state] (3) at (37.0,-5.325) {q3};
  \node[state] (4) at (36.975,-1.475) {q4};
  \node[state] (5) at (32.65,-1.425) {q5};
  \node[state] (6) at (29.35,-1.425) {q6};
  \node[state] (7) at (25.875,-1.425) {q7};
  \node[state] (8) at (22.3,-1.425) {q8};
  \node[state] (9) at (19.15,-1.425) {q9};
  \node[state] (10) at (16.125,-1.425) {q10};
  \node[state] (11) at (13.3,-1.425) {q11};
  \node[state] (12) at (10.25,-1.425) {q12};
  \node[state] (13) at (16.15,-3.625) {q13};
  \node[state] (14) at (7.35,-1.425) {q14};
  \node[state] (15) at (16.2,-5.775) {q15};
  \node[state,accepting] (16) at (20.45,-5.425) {NM};
  \node[state] (17) at (24.775,-4.65) {q17};
  \draw[->] (10) -- (11) node[midway, above] {\scriptsize 0, 1, L};
  \draw[->] (10) edge[loop above] node {\scriptsize 1, 1, R} (10);
  \draw[->] (4) -- (0) node[midway, above] {\scriptsize 0, 1, R};
  \draw[->] (4) edge[loop above] node {\scriptsize 1, 1, L} (4);
  \draw[->] (6) -- (17) node[midway, above] {\scriptsize 0, 0, R};
  \draw[->] (6) -- (7) node[midway, above] {\scriptsize 1, 0, R};
  \draw[->] (14) -- (15) node[midway, above] {\scriptsize 0, 0, L};
  \draw[->] (14) edge[loop above] node {\scriptsize 1, 1, L} (14);
  \draw[->] (7) -- (8) node[midway, above] {\scriptsize 0, 0, R};
  \draw[->] (7) edge[loop above] node {\scriptsize 1, 1, R} (7);
  \draw[->] (5) -- (6) node[midway, above] {\scriptsize 0, 0, R};
  \draw[->] (5) edge[loop above] node {\scriptsize 1, 1, L} (5);
  \draw[->] (3) -- (4) node[midway, above] {\scriptsize 0, 0, L};
  \draw[->] (3) edge[loop above] node {\scriptsize 1, 1, L} (3);
  \draw[->] (0) -- (5) node[midway, above] {\scriptsize 0, 0, L};
  \draw[->] (0) -- (1) node[midway, above] {\scriptsize 1, 0, R};
  \draw[->] (12) -- (13) node[midway, above] {\scriptsize 1, 1, L};
  \draw[->] (12) -- (14) node[midway, above] {\scriptsize 0, 1, L};
  \draw[->] (17) -- (16) node[midway, above] {\scriptsize 0, 1, R};
  \draw[->] (17) edge[loop above] node {\scriptsize 1, 0, R} (17);
  \draw[->] (11) -- (12) node[midway, above] {\scriptsize 0, 0, L};
  \draw[->] (11) edge[loop above] node {\scriptsize 1, 1, L} (11);
  \draw[->] (2) -- (3) node[midway, above] {\scriptsize 0, 1, L};
  \draw[->] (2) edge[loop above] node {\scriptsize 1, 1, R} (2);
  \draw[->] (15) -- (6) node[midway, above] {\scriptsize 0, 0, R};
  \draw[->] (15) edge[loop above] node {\scriptsize 1, 1, L} (15);
  \draw[->] (13) edge[loop above] node {\scriptsize 1, 1, L} (13);
  \draw[->] (13) -- (8) node[midway, above] {\scriptsize 0, 1, R};
  \draw[->] (9) -- (10) node[midway, above] {\scriptsize 0, 0, R};
  \draw[->] (9) edge[loop above] node {\scriptsize 1, 1, R} (9);
  \draw[->] (1) edge[loop above] node {\scriptsize 1, 1, R} (1);
  \draw[->] (1) -- (2) node[midway, above] {\scriptsize 0, 0, R};
  \draw[->] (8) -- (9) node[midway, above] {\scriptsize 1, 0, R};
\end{tikzpicture}
    \vspace{.3cm}
    C~~~FP Table 3 and Visualization
    \vspace{.5cm}
\setlength{\LTpre}{0pt}
\setlength{\LTpost}{0pt}
\begin{longtable}{|l|l|l|l|l|}
\hline
\textbf{State} & \textbf{Reads} & \textbf{Writes} & \textbf{Moves} & \textbf{Calls} \\
\hline
0 & 0 & 1 & R & 1 \\
\hline
0 & 1 & 1 & L & 0 \\
\hline
1 & 0 & 0 & R & 5 \\
\hline
1 & 1 & 0 & R & 2 \\
\hline
2 & 0 & 0 & R & 3 \\
\hline
2 & 1 & 1 & R & 2 \\
\hline
3 & 0 & 1 & L & 4 \\
\hline
3 & 1 & 1 & R & 3 \\
\hline
4 & 0 & 0 & L & 0 \\
\hline
4 & 1 & 1 & L & 4 \\
\hline
5 & 1 & 1 & R & 6 \\
\hline
6 & 1 & 1 & R & 7 \\
\hline
7 & 1 & 1 & R & 8 \\
\hline
8 & 1 & 0 & L & 9 \\
\hline
9 & 0 & 0 & R & 10 \\
\hline
9 & 1 & 1 & L & 9 \\
\hline
10 & 1 & 0 & L & 11 \\
\hline
11 & 1 & 0 & L & 12 \\
\hline
11 & 0 & 0 & L & 11 \\
\hline
12 & 1 & 1 & R & 13 \\
\hline
12 & 0 & 0 & R & 20 \\
\hline
13 & 1 & 0 & L & 14 \\
\hline
13 & 0 & 0 & R & 13 \\
\hline
14 & 0 & 1 & L & 15 \\
\hline
15 & 1 & 0 & L & 16 \\
\hline
15 & 0 & 0 & L & 15 \\
\hline
16 & 0 & 0 & R & 17 \\
\hline
16 & 1 & 1 & R & 32 \\
\hline
17 & 0 & 1 & R & 17 \\
\hline
17 & 1 & 1 & L & 40 \\
\hline
18 & 0 & 1 & R & 19 \\
\hline
18 & 1 & 1 & R & 18 \\
\hline
19 & 0 & 0 & R & 23 \\
\hline
19 & 1 & 1 & R & 19 \\
\hline
20 & 1 & 1 & L & 41 \\
\hline
20 & 0 & 1 & R & 20 \\
\hline
21 & 1 & 0 & L & 22 \\
\hline
22 & 0 & 0 & L & 15 \\
\hline
22 & 1 & 1 & L & 22 \\
\hline
23 & 1 & 0 & L & 30 \\
\hline
23 & 0 & 0 & R & 25 \\
\hline
24 & 1 & 1 & R & 43 \\
\hline
24 & 0 & 0 & R & 24 \\
\hline
26 & 1 & 1 & L & 31 \\
\hline
26 & 0 & 1 & R & 26 \\
\hline
27 & 1 & 1 & R & 26 \\
\hline
28 & 1 & 0 & L & 27 \\
\hline
28 & 0 & 0 & L & 28 \\
\hline
29 & 1 & 1 & L & 29 \\
\hline
29 & 0 & 0 & L & 28 \\
\hline
30 & 0 & 1 & L & 29 \\
\hline
31 & 1 & 0 & R & 10 \\
\hline
32 & 1 & 1 & R & 33 \\
\hline
32 & 0 & 0 & R & 32 \\
\hline
33 & 0 & 1 & R & 34 \\
\hline
33 & 1 & 1 & R & 33 \\
\hline
34 & 1 & 1 & R & 35 \\
\hline
35 & 0 & 0 & L & 36 \\
\hline
35 & 1 & 1 & L & 21 \\
\hline
36 & 0 & 0 & L & 37 \\
\hline
36 & 1 & 1 & L & 36 \\
\hline
37 & 1 & 0 & L & 38 \\
\hline
37 & 0 & 0 & L & 37 \\
\hline
38 & 1 & 1 & R & 39 \\
\hline
38 & 0 & 0 & R & 24 \\
\hline
39 & 1 & 0 & L & 11 \\
\hline
39 & 0 & 0 & R & 39 \\
\hline
40 & 1 & 0 & R & 18 \\
\hline
41 & 1 & 1 & L & 42 \\
\hline
42 & 1 & 0 & R & 19 \\
\hline
43 & 0 & 1 & L & 44 \\
\hline
43 & 1 & 1 & R & 43 \\
\hline
44 & 0 & 0 & R & 45 \\
\hline
44 & 1 & 1 & L & 44 \\
\hline
45 & 1 & 0 & R & NM \\
\hline

\end{longtable}

    \vspace{.3cm}
\begin{tikzpicture}[->,>=stealth,shorten >=1pt,auto,semithick,
    every node/.style={font=\tiny},
    xscale = .8,
    yscale = 1.4
]
\tikzstyle{every state}=[
    draw=black,
    thick,
    fill=white,
    minimum size=12pt,
    inner sep=1pt,
    font=\tiny
]
\node[state] (q0) at (20.1,-3.2) {  q0  };
\node[state, initial] (q1) at (16.7,1) {  q1  };
\node[state] (q2) at (17.7,1) {  q2  };
\node[state] (q3) at (18.9,1) {  q3  };
\node[state] (q4) at (20.1,1) {  q4  };
\node[state] (q5) at (16.7,-3.3) {  q5  };
\node[state] (q6) at (15.5,-3.3) {  q6  };
\node[state] (q7) at (15.5,-2.2) {  q7  };
\node[state] (q8) at (15.5,-1.1) {  q8  };
\node[state] (q9) at (15.5,-0.0) {  q9  };
\node[state] (q10) at (14.1,-0.0) {  q10  };
\node[state] (q11) at (12.6,-0.0) {  q11  };
\node[state] (q12) at (11.4,-0.0) {  q12  };
\node[state] (q13) at (10.0,-0.0) {  q13  };
\node[state] (q14) at (8.6,-0.0) {  q14  };
\node[state] (q15) at (7.2,-0.0) {  q15  };
\node[state] (q16) at (5.4,-0.6) {  q16  };
\node[state] (q17) at (1.6,-3) {  q17  };
\node[state] (q18) at (4.7,-3) {  q18  };
\node[state] (q19) at (5.9,-3) {  q19  };
\node[state] (q20) at (6.6,-1.8) {  q20  };
\node[state] (q21) at (0.0,-0.0) {  q21  };
\node[state] (q22) at (3.6,-0.0) {  q22  };
\node[state] (q23) at (2.85,-4.2) {  q23  };
\node[state] (q24) at (5.9,-6.5) {  q24  };
\node[state,accepting] (q25) at (1,-4.2) {  q25  };
\node[state] (q26) at (12.1,-4.2) {  q26  };
\node[state] (q27) at (10.25,-4.2) {  q27  };
\node[state] (q28) at (8.4,-4.2) {  q28  };
\node[state] (q29) at (6.55,-4.2) {  q29  };
\node[state] (q30) at (4.7,-4.2) {  q30  };
\node[state] (q31) at (14.1,-4.2) {  q31  };
\node[state] (q32) at (3.8,-1.1) {  q32  };
\node[state] (q33) at (2.5,-1.6) {  q33  };
\node[state] (q34) at (1.2,-2.0) {  q34  };
\node[state] (q35) at (0.0,-2.4) {  q35  };
\node[state] (q36) at (0.0,-5.3) {  q36  };
\node[state] (q37) at (3.4,-5.3) {  q37  };
\node[state] (q38) at (7.2,-5.3) {  q38  };
\node[state] (q39) at (12.6,-5.3) {  q39  };
\node[state] (q40) at (3.2,-3) {  q40  };
\node[state] (q41) at (8.4,-1.8) {  q41  };
\node[state] (q42) at (9.7,-1.8) {  q42  };
\node[state] (q43) at (4.5,-6.5) {  q43  };
\node[state] (q44) at (3.0,-6.5) {  q44  };
\node[state] (q45) at (1.5,-6.5) {  q45  };
\node[state] (NM) at (0.0,-6.5) {  NM  };

\path (q2) edge node { 0,0,R  } (q3);
\path (q35) edge node { 0,0,L  } (q36);
\path (q41) edge node {  1,1,L  } (q42);
\path (q28) edge node {  1,0,L  } (q27);
\path (q18) edge node { 0,1,R  } (q19);
\path (q14) edge node { 0,1,L  } (q15);
\path (q6) edge node {  1,1,R  } (q7);
\path (q4) edge node { 0,0,L  } (q0);
\path (q16) edge node { 0,0,R  } (q17);
\path (q9) edge node { 0,0,R  } (q10);
\path (q23) edge node {  1,0,L  } (q30);
\path (q8) edge node {  1,0,L  } (q9);
\path (q21) edge node {  1,0,L  } (q22);
\path (q10) edge node {  1,0,L  } (q11);
\path (q36) edge node { 0,0,L  } (q37);
\path (q23) edge node { 0,0,R  } (q25);
\path (q33) edge node { 0,1,R  } (q34);
\path (q13) edge node {  1,0,L  } (q14);
\path (q26) edge node {  1,1,L  } (q31);
\path (q5) edge node {  1,1,R  } (q6);
\path (q20) edge node {  1,1,L  } (q41);
\path (q39) edge node {  1,0,L  } (q11);
\path (q12) edge node {  1,1,R  } (q13);
\path (q38) edge node {  1,1,R  } (q39);
\path (q7) edge node {  1,1,R  } (q8);
\path (q22) edge node { 0,0,L  } (q15);
\path (q1) edge node { 0,0,R  } (q5);
\path (q34) edge node {  1,1,R  } (q35);
\path (q0) edge node { 0,1,R  } (q1);
\path (q45) edge node {  1,0,R  } (NM);
\path (q3) edge node { 0,1,L  } (q4);
\path (q16) edge node {  1,1,R  } (q32);
\path (q27) edge node {  1,1,R  } (q26);
\path (q19) edge node { 0,0,R  } (q23);
\path (q24) edge node {  1,1,R  } (q43);
\path (q32) edge node {  1,1,R  } (q33);
\path (q37) edge node {  1,0,L  } (q38);
\path (q42) edge node {  1,0,R  } (q19);
\path (q30) edge node { 0,1,L  } (q29);
\path (q43) edge node { 0,1,L  } (q44);
\path (q26) edge [loop above] node { 0,1,R  } (q26);
\path (q17) edge [loop above] node { 0,1,R  } (q17);
\path (q20) edge [loop above] node { 0,1,R  } (q20);
\path (q17) edge node {  1,1,L  } (q40);
\path (q35) edge node {  1,1,L  } (q21);
\path (q15) edge node {  1,0,L  } (q16);
\path (q11) edge node {  1,0,L  } (q12);
\path (q44) edge node { 0,0,R  } (q45);
\path (q38) edge node { 0,0,R  } (q24);
\path (q1) edge node {  1,0,R  } (q2);
\path (q36) edge [loop above] node {  1,1,L  } (q36);
\path (q0) edge [loop above] node {  1,1,L  } (q0);
\path (q15) edge [loop above] node { 0,0,L  } (q15);
\path (q28) edge [loop above] node { 0,0,L  } (q28);
\path (q29) edge [loop above] node {  1,1,L  } (q29);
\path (q9) edge [loop above] node {  1,1,L  } (q9);
\path (q37) edge [loop above] node { 0,0,L  } (q37);
\path (q44) edge [loop above] node {  1,1,L  } (q44);
\path (q4) edge [loop above] node {  1,1,L  } (q4);
\path (q22) edge [loop above] node {  1,1,L  } (q22);
\path (q11) edge [loop above] node { 0,0,L  } (q11);
\path (q31) edge node {  1,0,R  } (q10);
\path (q39) edge [loop above] node { 0,0,R  } (q39);
\path (q18) edge [loop above] node {  1,1,R  } (q18);
\path (q2) edge [loop above] node {  1,1,R  } (q2);
\path (q32) edge [loop above] node { 0,0,R  } (q32);
\path (q13) edge [loop above] node { 0,0,R  } (q13);
\path (q33) edge [loop above] node {  1,1,R  } (q33);
\path (q3) edge [loop above] node {  1,1,R  } (q3);
\path (q19) edge [loop above] node {  1,1,R  } (q19);
\path (q43) edge [loop above] node {  1,1,R  } (q43);
\path (q24) edge [loop above] node { 0,0,R  } (q24);
\path (q40) edge node {  1,0,R  } (q18);
\path (q12) edge node { 0,0,R  } (q20);
\path (q29) edge node { 0,0,L  } (q28);

\end{tikzpicture}
    \vspace{.3cm}
    D~~~BP Table and Visualization
    \vspace{.5cm}
\setlength{\LTpre}{0pt}
\setlength{\LTpost}{0pt}
\begin{longtable}{|l|l|l|l|l|}
\hline
\textbf{State} & \textbf{Reads} & \textbf{Writes} & \textbf{Moves} & \textbf{Calls} \\
\hline
Af & 0 & 1 & R & Bf \\
\hline
Af & 1 & 1 & L & Af \\
\hline
Gf & 0 & 0 & L & Hf \\
\hline
Gf & 1 & 1 & L & Gf \\
\hline
Hf & 1 & 1 & L & If \\
\hline
Hf & 0 & 1 & L & Kf \\
\hline
If & 0 & 1 & R & Jf \\
\hline
If & 1 & 1 & L & If \\
\hline
Cf & 0 & 0 & R & Df \\
\hline
Cf & 1 & 1 & R & Cf \\
\hline
Jf & 1 & 0 & R & Ef \\
\hline
Kf & 0 & 0 & L & Lf \\
\hline
Kf & 1 & 1 & L & Kf \\
\hline
Lf & 0 & 1 & L & Mf \\
\hline
Lf & 1 & 1 & L & Af \\
\hline
Pf & 0 & 0 & R & Vf \\
\hline
Pf & 1 & 1 & L & Qf \\
\hline
Ff & 0 & 1 & L & Gf \\
\hline
Ff & 1 & 1 & R & Ff \\
\hline
Qf & 1 & 1 & L & Rf \\
\hline
Qf & 0 & 0 & R & Vf \\
\hline
Mf & 1 & 1 & L & Nf \\
\hline
Nf & 1 & 1 & L & Of \\
\hline
Of & 1 & 1 & L & Pf \\
\hline
Of & 0 & 0 & R & Vf \\
\hline
Rf & 0 & 0 & R & Vf \\
\hline
Rf & 1 & 1 & L & Sf \\
\hline
Sf & 1 & 1 & R & Tf \\
\hline
Sf & 0 & 0 & R & Vf \\
\hline
Df & 1 & 0 & R & Ef \\
\hline
Tf & 0 & 0 & R & Uf \\
\hline
Tf & 1 & 1 & R & Tf \\
\hline
Ef & 0 & 0 & R & Ff \\
\hline
Ef & 1 & 1 & R & Ef \\
\hline
Bf & 1 & 0 & R & Cf \\
\hline
Uf & 0 & 1 & R & As \\
\hline
Uf & 1 & 1 & R & Uf \\
\hline
Vf & 0 & 0 & R & Wf \\
\hline
Vf & 1 & 1 & R & Vf \\
\hline
Wf & 0 & 1 & R & Xf \\
\hline
Wf & 1 & 1 & R & Wf \\
\hline
Rs & 1 & 0 & R & Qs \\
\hline
Rs & 0 & 0 & L & Ss \\
\hline
Qs & 0 & 0 & R & Rs \\
\hline
Qs & 1 & 0 & R & Qs \\
\hline
Xf & 1 & 1 & R & Xf \\
\hline
Xf & 0 & 0 & L & Ss \\
\hline
Ps & 0 & 0 & R & Qs \\
\hline
Os & 1 & 0 & R & Ps \\
\hline
Os & 0 & 1 & R & Os \\
\hline
Ns & 0 & 1 & L & HALT \\
\hline
Ns & 1 & 1 & R & Ks \\
\hline
Ms & 1 & 0 & L & Ns \\
\hline
Ms & 0 & 0 & L & Ms \\
\hline
Ls & 0 & 0 & L & Ls \\
\hline
Ls & 1 & 1 & L & Ms \\
\hline
Ks & 1 & 1 & R & Bs \\
\hline
Ks & 0 & 0 & R & Ks \\
\hline
Js & 0 & 0 & R & Os \\
\hline
Js & 1 & 1 & R & Ks \\
\hline
Is & 1 & 0 & L & Js \\
\hline
Is & 0 & 0 & R & Os \\
\hline
Hs & 1 & 0 & L & Is \\
\hline
Hs & 0 & 0 & L & Hs \\
\hline
Gs & 1 & 1 & L & Hs \\
\hline
Fs & 0 & 0 & L & Gs \\
\hline
Fs & 1 & 1 & L & Fs \\
\hline
Es & 1 & 0 & L & Fs \\
\hline
Ds & 0 & 1 & L & Es \\
\hline
Ds & 1 & 1 & R & Ds \\
\hline
Cs & 1 & 1 & R & Ds \\
\hline
Cs & 0 & 1 & L & Ls \\
\hline
Bs & 0 & 0 & R & Cs \\
\hline
As & 0 & 1 & R & Bs \\
\hline
As & 0 & 1 & R & As \\
\hline
Ss & 0 & 1 & L & Kf \\
\hline
Ss & 0 & 0 & L & Ss \\
\hline
Df & 0 & 1 & L & Jf \\
\hline
Jf & 0 & 1 & L & Pf \\
\hline
Pf & 0 & 0 & L & Qf \\
\hline
Qf & 1 & 0 & L & Es \\
\hline
Es & 0 & 1 & L & Af \\
\hline
\end{longtable}
    \vspace{.3cm}
\begin{tikzpicture}[->,>=stealth,shorten >=1pt,auto,semithick,
    every node/.style={font=\tiny},
    xscale=.45,
    yscale=.5
]
\tikzstyle{every state}=[
    draw=black,
    thick,
    fill=white,
    minimum size=12pt,
    inner sep=1pt,
    font=\tiny
]
  \node[state] (0) at (36.175,-13.225) {Af};
  \node[state] (1) at (25.275,-10.675) {Gf};
  \node[state] (2) at (28.875,-10.675) {Hf};
  \node[state] (3) at (33.1,-10.675) {If};
  \node[state] (4) at (36.225,-2.625) {Cf};
  \node[state] (5) at (33.1,-5.975) {Jf};
  \node[state] (6) at (28.875,-16.1) {Kf};
  \node[state] (7) at (36.175,-16.125) {Lf};
  \node[state] (8) at (41.275,-20.6) {Pf};
  \node[state] (9) at (25.275,-6.2) {Ff};
  \node[state] (10) at (41.25,-22.85) {Qf};
  \node[state] (11) at (38.9,-16.15) {Mf};
  \node[state] (12) at (41.275,-16.15) {Nf};
  \node[state] (13) at (41.25,-18.25) {Of};
  \node[state] (14) at (41.25,-25.2) {Rf};
  \node[state] (15) at (41.25,-27.575) {Sf};
  \node[state] (16) at (33.125,-2.6) {Df};
  \node[state] (17) at (31.0,-27.575) {Tf};
  \node[state] (18) at (28.9,-2.6) {Ef};
  \node[state] (19) at (36.175,-8.775) {Bf};
  \node[state] (20) at (28.2,-27.575) {Uf};
  \node[state] (21) at (36.95,-22.35) {Vf};
  \node[state] (22) at (33.425,-22.4) {Wf};
  \node[state] (23) at (25.025,-14.55) {Rs};
  \node[state] (24) at (21.8,-14.575) {Qs};
  \node[state] (25) at (30.05,-22.4) {Xf};
  \node[state] (26) at (18.425,-12.1) {Ps};
  \node[state] (27) at (14.7,-12.075) {Os};
  \node[state] (28) at (15.175,-21.7) {Ns};
  \node[state] (29) at (12.25,-21.7) {Ms};
  \node[state] (30) at (9.25,-21.7) {Ls};
  \node[state] (31) at (19.975,-21.725) {Ks};
  \node[state] (32) at (19.95,-15.325) {Js};
  \node[state] (33) at (11.675,-15.35) {Is};
  \node[state] (34) at (8.725,-15.325) {Hs};
  \node[state] (35) at (5.85,-15.325) {Gs};
  \node[state] (36) at (5.95,-19.175) {Fs};
  \node[state] (37) at (5.95,-23.225) {Es};
  \node[state] (38) at (5.95,-27.55) {Ds};
  \node[state] (39) at (9.25,-27.55) {Cs};
  \node[state] (40) at (19.975,-27.575) {Bs};
  \node[state] (41) at (24.175,-27.575) {As};
  \node[state,accepting] (42) at (15.175,-24.5) {HALT};
  \node[state] (43) at (25.025,-22.375) {Ss};
  \node[state,initial] (44) at (39.025,-2.625) {Df};
  \node[state] (45) at (39.025,-5.2) {Jf};
  \node[state] (46) at (39.025,-7.925) {Pf};
  \node[state] (47) at (39.025,-10.75) {Qf};
  \node[state] (48) at (39.025,-13.225) {Es};
  \draw[->] (2) -- (3) node[midway, above] {\scriptsize 1, 1, L};
  \draw[->] (2) -- (6) node[midway, above] {\scriptsize 0, 1, L};
  \draw[->] (8) -- (21) node[midway, above] {\scriptsize 0, 0, R};
  \draw[->] (8) -- (10) node[midway, above] {\scriptsize 1, 1, L};
  \draw[->] (31) -- (40) node[midway, above] {\scriptsize 1, 1, R};
  \draw[->] (31) edge[loop above] node {\scriptsize 0, 0, R} (31);
  \draw[->] (10) -- (14) node[midway, above] {\scriptsize 1, 1, L};
  \draw[->] (10) -- (21) node[midway, above] {\scriptsize 0, 0, R};
  \draw[->] (43) -- (6) node[midway, above] {\scriptsize 0, 1, L};
  \draw[->] (43) edge[loop above] node {\scriptsize 0, 0, L} (43);
  \draw[->] (15) -- (17) node[midway, above] {\scriptsize 1, 1, R};
  \draw[->] (15) -- (21) node[midway, above] {\scriptsize 0, 0, R};
  \draw[->] (21) -- (22) node[midway, above] {\scriptsize 0, 0, R};
  \draw[->] (21) edge[loop above] node {\scriptsize 1, 1, R} (21);
  \draw[->] (5) -- (18) node[midway, above] {\scriptsize 1, 0, R};
  \draw[->] (20) -- (41) node[midway, above] {\scriptsize 0, 1, R};
  \draw[->] (20) edge[loop above] node {\scriptsize 1, 1, R} (20);
  \draw[->] (14) -- (21) node[midway, above] {\scriptsize 0, 0, R};
  \draw[->] (14) -- (15) node[midway, above] {\scriptsize 1, 1, L};
  \draw[->] (29) -- (28) node[midway, above] {\scriptsize 1, 0, L};
  \draw[->] (29) edge[loop above] node {\scriptsize 0, 0, L} (29);
  \draw[->] (26) -- (24) node[midway, above] {\scriptsize 0, 0, R};
  \draw[->] (28) -- (42) node[midway, above] {\scriptsize 0, 1, L};
  \draw[->] (28) -- (31) node[midway, above] {\scriptsize 1, 1, R};
  \draw[->] (36) -- (35) node[midway, above] {\scriptsize 0, 0, L};
  \draw[->] (36) edge[loop above] node {\scriptsize 1, 1, L} (36);
  \draw[->] (17) -- (20) node[midway, above] {\scriptsize 0, 0, R};
  \draw[->] (17) edge[loop above] node {\scriptsize 1, 1, R} (17);
  \draw[->] (11) -- (12) node[midway, above] {\scriptsize 1, 1, L};
  \draw[->] (41) -- (40) node[midway, above] {\scriptsize 0, 1, R};
  \draw[->] (41) edge[loop above] node {\scriptsize 0, 1, R} (41);
  \draw[->] (33) -- (32) node[midway, above] {\scriptsize 1, 0, L};
  \draw[->] (33) -- (27) node[midway, above] {\scriptsize 0, 0, R};
  \draw[->] (6) -- (7) node[midway, above] {\scriptsize 0, 0, L};
  \draw[->] (6) edge[loop above] node {\scriptsize 1, 1, L} (6);
  \draw[->] (48) -- (0) node[midway, above] {\scriptsize 0, 1, L};
  \draw[->] (13) -- (8) node[midway, above] {\scriptsize 1, 1, L};
  \draw[->] (13) -- (21) node[midway, above] {\scriptsize 0, 0, R};
  \draw[->] (32) -- (27) node[midway, above] {\scriptsize 0, 0, R};
  \draw[->] (32) -- (31) node[midway, above] {\scriptsize 1, 1, R};
  \draw[->] (16) -- (18) node[midway, above] {\scriptsize 1, 0, R};
  \draw[->] (47) -- (48) node[midway, above] {\scriptsize 1, 0, L};
  \draw[->] (18) -- (9) node[midway, above] {\scriptsize 0, 0, R};
  \draw[->] (18) edge[loop above] node {\scriptsize 1, 1, R} (18);
  \draw[->] (37) -- (36) node[midway, above] {\scriptsize 1, 0, L};
  \draw[->] (9) -- (1) node[midway, above] {\scriptsize 0, 1, L};
  \draw[->] (9) edge[loop above] node {\scriptsize 1, 1, R} (9);
  \draw[->] (4) -- (16) node[midway, above] {\scriptsize 0, 0, R};
  \draw[->] (4) edge[loop above] node {\scriptsize 1, 1, R} (4);
  \draw[->] (45) -- (46) node[midway, above] {\scriptsize 0, 1, L};
  \draw[->] (27) -- (26) node[midway, above] {\scriptsize 1, 0, R};
  \draw[->] (27) edge[loop above] node {\scriptsize 0, 1, R} (27);
  \draw[->] (38) -- (37) node[midway, above] {\scriptsize 0, 1, L};
  \draw[->] (38) edge[loop above] node {\scriptsize 1, 1, R} (38);
  \draw[->] (39) -- (38) node[midway, above] {\scriptsize 1, 1, R};
  \draw[->] (39) -- (30) node[midway, above] {\scriptsize 0, 1, L};
  \draw[->] (23) -- (24) node[midway, above] {\scriptsize 1, 0, R};
  \draw[->] (23) -- (43) node[midway, above] {\scriptsize 0, 0, L};
  \draw[->] (24) -- (23) node[midway, above] {\scriptsize 0, 0, R};
  \draw[->] (24) edge[loop above] node {\scriptsize 1, 0, R} (24);
  \draw[->] (0) -- (19) node[midway, above] {\scriptsize 0, 1, R};
  \draw[->] (0) edge[loop above] node {\scriptsize 1, 1, L} (0);
  \draw[->] (19) -- (4) node[midway, above] {\scriptsize 1, 0, R};
  \draw[->] (3) -- (5) node[midway, above] {\scriptsize 0, 1, R};
  \draw[->] (3) edge[loop above] node {\scriptsize 1, 1, L} (3);
  \draw[->] (35) -- (34) node[midway, above] {\scriptsize 1, 1, L};
  \draw[->] (1) -- (2) node[midway, above] {\scriptsize 0, 0, L};
  \draw[->] (1) edge[loop above] node {\scriptsize 1, 1, L} (1);
  \draw[->] (34) -- (33) node[midway, above] {\scriptsize 1, 0, L};
  \draw[->] (34) edge[loop above] node {\scriptsize 0, 0, L} (34);
  \draw[->] (22) -- (25) node[midway, above] {\scriptsize 0, 1, R};
  \draw[->] (22) edge[loop above] node {\scriptsize 1, 1, R} (22);
  \draw[->] (12) -- (13) node[midway, above] {\scriptsize 1, 1, L};
  \draw[->] (7) -- (11) node[midway, above] {\scriptsize 0, 1, L};
  \draw[->] (7) -- (0) node[midway, above] {\scriptsize 1, 1, L};
  \draw[->] (46) -- (47) node[midway, above] {\scriptsize 0, 0, L};
  \draw[->] (30) edge[loop above] node {\scriptsize 0, 0, L} (30);
  \draw[->] (30) -- (29) node[midway, above] {\scriptsize 1, 1, L};
  \draw[->] (40) -- (39) node[midway, above] {\scriptsize 0, 0, R};
  \draw[->] (25) edge[loop above] node {\scriptsize 1, 1, R} (25);
  \draw[->] (25) -- (43) node[midway, above] {\scriptsize 0, 0, L};
  \draw[->] (44) -- (45) node[midway, above] {\scriptsize 0, 1, L};
\end{tikzpicture}
}\end{flushleft}

\vfill

\nocite{*}
}\end{flushleft}
\end{document}